\documentclass[longauth]{aa}  
\pdfoutput=0
\usepackage{txfonts}
\usepackage{textcomp}
\usepackage{graphicx}
\usepackage{epstopdf}
\usepackage{longtable}
\usepackage{color}
\usepackage{multirow}
\usepackage[numbers]{natbib}
\bibpunct{[}{]}{,}{n}{}{;}

\begin{document}

\title{First detection of VHE $\gamma$--rays from SN~1006 by H.E.S.S.}
\offprints{naumann-godo@llr.in2p3.fr, denauroi@in2p3.fr}

\small{
\author{ HESS Collaboration
 \and F.~Acero \inst{15}
 \and F. Aharonian\inst{1,13}
 \and A.G.~Akhperjanian \inst{2}
 \and G.~Anton \inst{16}
 \and U.~Barres de Almeida \inst{8} \thanks{supported by CAPES Foundation, Ministry of Education of Brazil}
 \and A.R.~Bazer-Bachi \inst{3}
 \and Y.~Becherini \inst{12}
 \and B.~Behera \inst{14}
 \and M.~Beilicke \inst{4} \thanks{now at Washington University, St. Louis, USA}
 \and K.~Bernl\"ohr \inst{1,5}
% ...
}
}

\newpage

\institute{
Max-Planck-Institut f\"ur Kernphysik, P.O. Box 103980, D 69029
Heidelberg, Germany
\and
 Yerevan Physics Institute, 2 Alikhanian Brothers St., 375036 Yerevan,
Armenia
\and
Centre d'Etude Spatiale des Rayonnements, CNRS/UPS, 9 av. du Colonel Roche, BP
4346, F-31029 Toulouse Cedex 4, France
\and
Universit\"at Hamburg, Institut f\"ur Experimentalphysik, Luruper Chaussee
149, D 22761 Hamburg, Germany
\and
Institut f\"ur Physik, Humboldt-Universit\"at zu Berlin, Newtonstr. 15,
D 12489 Berlin, Germany
\and
LUTH, Observatoire de Paris, CNRS, Universit\'e Paris Diderot, 5 Place Jules Janssen, 92190 Meudon, 
France
\and
IRFU/DSM/CEA, CE Saclay, F-91191
Gif-sur-Yvette, Cedex, France
\and
University of Durham, Department of Physics, South Road, Durham DH1 3LE,
U.K.
\and
Unit for Space Physics, North-West University, Potchefstroom 2520,
    South Africa
\and
Laboratoire Leprince-Ringuet, Ecole Polytechnique, CNRS/IN2P3,
 F-91128 Palaiseau, France
\and 
Laboratoire d'Annecy-le-Vieux de Physique des Particules,
Universit\'{e} de Savoie, CNRS/IN2P3, F-74941 Annecy-le-Vieux,
France
\and
Astroparticule et Cosmologie (APC), CNRS, Universite Paris 7 Denis Diderot,
10, rue Alice Domon et Leonie Duquet, F-75205 Paris Cedex 13, France
\thanks{UMR 7164 (CNRS, Universit\'e Paris VII, CEA, Observatoire de Paris)}
\and
Dublin Institute for Advanced Studies, 5 Merrion Square, Dublin 2,
Ireland
\and
Landessternwarte, Universit\"at Heidelberg, K\"onigstuhl, D 69117 Heidelberg, Germany
\and
Laboratoire de Physique Th\'eorique et Astroparticules, 
Universit\'e Montpellier 2, CNRS/IN2P3, CC 70, Place Eug\`ene Bataillon, F-34095
Montpellier Cedex 5, France
\and
Universit\"at Erlangen-N\"urnberg, Physikalisches Institut, Erwin-Rommel-Str. 1,
D 91058 Erlangen, Germany
% ...
}

\date{Released 2009 Xxxxx XX}

\abstract
{}
{Recent theoretical predictions of the lowest very high energy (VHE) luminosity of SN~1006 are only a factor 5 below the previously
published H.E.S.S. upper limit, thus motivating further in-depth
observations of this source.}
{Deep observations at VHE energies (above 100~GeV) were carried out with the High Energy Stereoscopic System (H.E.S.S.) of Cherenkov Telescopes from 2003 to 2008. More than 100 hours of data have been collected and subjected to an improved analysis procedure. }
{Observations resulted in the detection of VHE $\gamma$-rays from SN~1006. 
The measured $\gamma$-ray spectrum is compatible with a power-law, the flux is of the order of 1$\%$ of that detected from the Crab Nebula, and is thus consistent with the previously established H.E.S.S.\ upper limit. The source exhibits a bipolar morphology, which is strongly correlated with non-thermal X-rays. }
{Because the thickness of the VHE-shell is compatible with emission from
a thin rim, particle acceleration in shock waves is likely to be the
origin of the $\gamma$-ray signal.  The measured flux level can be
accounted for by inverse Compton emission, but a mixed scenario that includes
leptonic and hadronic components and takes into account the ambient
matter density inferred from observations also leads to a satisfactory
description of the multi-wavelength spectrum.}

\maketitle
%

\section{Introduction}\label{intro}

The source SN~1006 is the remnant of one of the few historical supernovae.
It appeared in the southern sky on 1006 May 1 and was recorded by Chinese and Arab astronomers \cite{Stephenson02}.
The remnant of this explosion was first identified at radio wavelengths on the basis of historical 
evidence \cite{Gardner}. The evolution of its luminosity indicates that it is the result of a Type Ia supernova \cite{Schaefer}, probably 
the brightest supernova in recorded history. A distance of 2.2~kpc was derived by Winkler et~al.~(2003) based on comparing the optical proper motion with an estimate of the shock velocity derived from optical thermal line broadening assuming a high Mach number single-fluid shock. 

Contemporary interest in the very high energy (VHE) emission from supernova remnants (SNRs) has arisen due to their association as prime candidates for Galactic cosmic-ray acceleration. Firstly, Galactic SNRs have sufficient kinetic energy to explain the estimated Galactic luminosity 
in cosmic rays of $10^{40}$~erg/s. Secondly, and more importantly, it has been shown that diffusive shock acceleration provides a viable mechanism 
which can efficiently accelerate charged particles in the blast waves of SNRs (e.g. Drury 1983; Blandford \& Eichler 1987; Jones \& Ellison 1991; Berezhko et al. 1996). 
Indeed, most shell-type SNRs are non-thermal radio emitters, which confirms that electrons are accelerated up to at least GeV energies. Moreover, the limb-brightened non-thermal radio emission traces the site of effective particle acceleration. 

The source SN~1006 was also the first SNR in which a non-thermal component of hard X-rays was detected in the rims of the remnant by ASCA \cite{Koyama} and ROSAT \cite{Willingale}, whereas the interior of the remnant exhibits a thermal spectrum with line emission. The hard featureless power-law spectrum strongly implies a synchrotron origin of the radiation, which in turn suggests that electrons can be accelerated up to energies of $\sim\mathrm{100}$~TeV. 
Subsequent arcsecond resolution images by Chandra revealed a small-scale structure in the nonthermal X-ray filaments of the NE rim of 
SN~1006 \cite{Bamba,Long}, supporting the idea of high B-fields in the bright limbs of the remnant \cite{Berezhko2002}. 
An analysis of the X-ray observations from XMM-Newton by Rothenflug et.~al (2004) leads to the conclusion that the magnetic field in the remnant is oriented in the NE-SW direction. The synchrotron emission would then be concentrated in regions where the shock is quasi-parallel \cite{Voelk2003}. 

Also, $\gamma$-ray observations of SN~1006 were carried out by ground-based $\gamma$-ray telescopes. A TeV $\gamma$-ray signal at the level of the Crab flux was claimed by the CANGAROO-I \cite{Cangaroo-I} and CANGAROO-II \cite{Cangaroo-II} telescopes, but subsequent stereoscopic observations of the source with the H.E.S.S. telescopes in 2003 and 2004 found no evidence of VHE $\gamma$-ray emission and derived an upper limit of $\Phi(>0.26$~TeV$)<2.4\times 10^{-12}$~ph~cm$^{-2}$~s$^{-1}$ at 99.9\% confidence level \cite{HESS_limit}. The CANGAROO-III telescope array found only an upper limit which is consistent with the H.E.S.S. result \cite{Cangaroo-III}. 

The initial non-detection of SN~1006 in VHE $\gamma$-rays does not invalidate the hypothesis of nuclear particle acceleration in the shock. Indeed, the hadronic $\gamma$-ray flux is very sensitive to the ambient gas density $\mathrm{n_H}$ and hence the H.E.S.S. upper limit implies a constraint on $\mathrm{n_H}<0.1$~cm$^{-3}$ \cite{Ksenofontov}. Indeed, being 500~pc above the Galactic plane, the remnant is relatively isolated, and the gas density around SN~1006 was 
recently estimated to be around 0.085~cm$^{-3}$ \cite{Katsuda}.
Ksenofontov et al.~(2005) furthermore showed that the lower limit for the VHE $\gamma$-ray flux, which is given by the inverse Compton (IC) component derived from the integrated synchrotron flux and field amplification alone, was only a factor 5 below the H.E.S.S. upper limit. 
These predictions promoted deep observations with the H.E.S.S. telescopes. 



\section{H.E.S.S. observations and analysis methods}\label{analysis}

H.E.S.S. is an array of four 13~m diameter imaging atmospheric Cherenkov telescopes situated in the Khomas Highland in Namibia at an altitude 
of 1800~m above sea level \cite{Bernloehr,Funk}. The source SN~1006 was observed in 2003 with the two telescopes that were operational at that time and with
the complete H.E.S.S.~array in the years since. 
After run selection the data set comprises 130 hours (live time) of observations, of which 18 hours were taken with two telescopes only. The latter yielded 
a smaller effective area than the data set recorded with the full array. For that reason they are used only in morphological studies and excluded in the spectral analysis.

The data were analysed with the {\it Model Analysis} \cite{Naurois}, in which shower images of all triggered telescopes are compared to a pre-calculated model by means of a log-likelihood minimisation. The Model Analysis does not rely on any image-cleaning procedure and uses all pixels in the camera. The noise distributions in the pixels due to the night sky background are taken into account in the model fit and result in a superior treatment of shower tails. Therefore the Model Analysis results in a more precise reconstruction and better background suppression  than  more conventional techniques, thus leading to improved sensitivity.

Two different sets of cuts were used: The {\it standard cuts}, including a minimum image charge
of 60 photoelectrons ($\mathrm{E_{th}} = 260$~GeV), cover the full energy range and are used for the spectral analysis only. The {\it hard cuts}, with a larger charge 
cut of 200 photoelectrons, result in an improved signal-to-background ratio at the expense of lower statistics and a higher threshold of 500~GeV. These are used for the studies of source morphology.

The results presented below have been cross-checked using the 3D Model Analysis \cite{Marianne,Melitta}. Both analyses yielded consistent results.

\begin{figure}
\begin{center}
\includegraphics[width=0.95\linewidth]{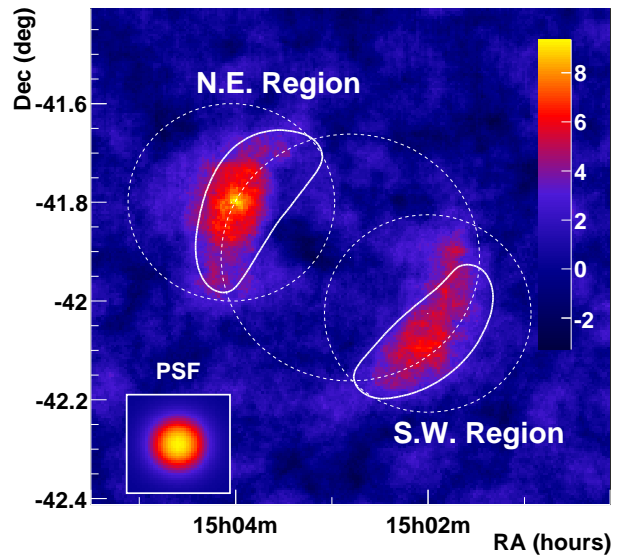}
\caption{\it H.E.S.S. $\gamma$-ray significance map of SN~1006 using an integration radius of $0.05^\circ$. The linear colour scale is in units of standard deviations. The white solid contours correspond to the regions which contain 80\% of the non-thermal X-ray emission from the XMM-Newton flux map in the 2 - 4.5~keV energy range after smearing with the H.E.S.S. PSF, shown in the inset. The white dashed circles correspond to the regions that are excluded from background determination.}
\label{fig:significance200}
\end{center}
\end{figure}

Significant $\gamma$-ray emission is detected from the direction of SN~1006, concentrated in two extended regions as shown in Fig. \ref{fig:significance200}. This map shows the significance over a field-of-view of $1^\circ \times 1^\circ$ with a pixel size of $0.005^\circ$ obtained with hard cuts using the ring background subtraction technique \cite{Berge} and a small integration radius of $0.05^\circ$, close to the H.E.S.S. PSF of $R_{68}=0.064^\circ$. As the pixel size is a factor 10 smaller than the integration radius, the bins are highly correlated. In two regions of the map the significance of the H.E.S.S. observation clearly exceeds 5~$\sigma$.

\begin{figure}
\begin{center}
\includegraphics[width=0.95\linewidth]{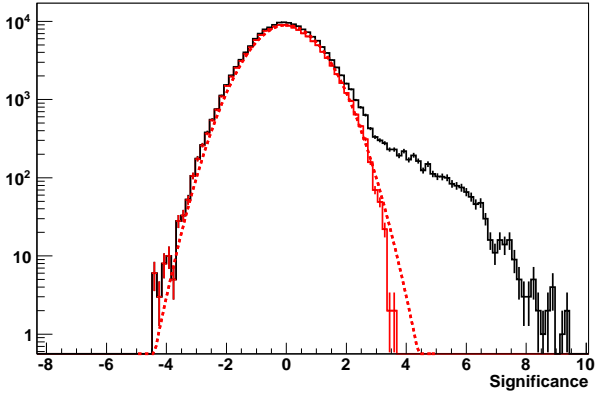}
\caption{\it H.E.S.S. $\gamma$-ray significance distribution over the full field-of-view of SN~1006 (black histogram) and excluding the circular regions around the NE and SW emission regions (red histogram). A normal distribution (red dashed line) shows that the significance distribution over the rest of the field-of-view is compatible with expectation from statistical noise fluctuations.}
\label{fig:sigdist200}
\end{center}
\end{figure}

\begin{figure}
\begin{center}
\includegraphics[width=0.95\linewidth]{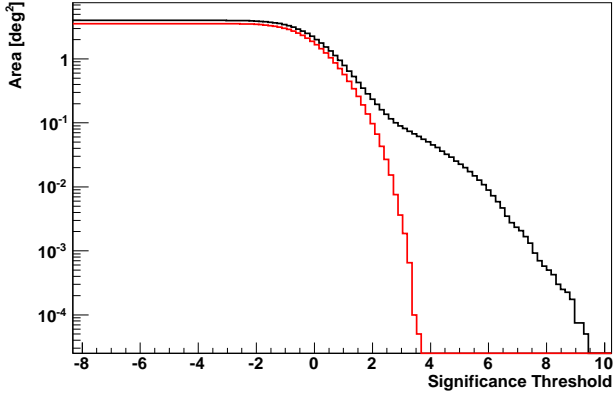}
\caption{\it H.E.S.S. sky area with $\gamma$-ray significance above some threshold as a function of its value over the full field-of-view of SN~1006 (black histogram) and excluding the circular regions around the NE and SW emission regions (red histogram). }
\label{fig:cumulsigdist200}
\end{center}
\end{figure}

The significance distribution over the field-of-view of $2^\circ \times 2^\circ$ is shown in Fig.~\ref{fig:sigdist200}, while Fig.~\ref{fig:cumulsigdist200} illustrates the area corresponding to the significance above a given level. 
The black histogram in both figures corresponds to the full field-of-view and exhibits strong deviation from a normal distribution at large significance values. The red histogram, restricted to the part of the field-of-view outside of the white dashed circles (Fig.~\ref{fig:significance200}) is compatible with a normal distribution, as denoted by the red dashed line (Fig.~\ref{fig:sigdist200}). This demonstrates that the distribution of events over the field-of-view (outside the two exclusion regions) is compatible with expectation from statistical fluctuations and that systematic effects concerning background estimation are under control. 


\section{Morphology}\label{morphology}

Two different integration regions were defined {\it a priori} from the XMM-Newton data set \cite{Rothenflug}: a map of the flux in the 2 - 4.5 keV energy range (to exclude thermal contamination) was smoothed with the H.E.S.S. PSF, and regions which contained $80\%$ of the flux were calculated. The two resulting regions, denoted as {\it NE Region} and {\it SW Region}, are displayed as white contours in Fig. \ref{fig:significance200} and coincide well with the regions of largest H.E.S.S. significance.

Excess event counts and significances for both regions are given in Table \ref{tab:sig} for the two sets of cuts. 
The ON photons are from the regions enclosed by the solid lines in Fig.~\ref{fig:significance200}, while the OFF events are
taken from regions of identical shape rotated in the field-of-view of the instrument around
the observation position and not intersecting the exclusion regions (enclosed by dashed
lines in Fig.~\ref{fig:significance200}). 
Due to varying observation positions, the number of OFF regions varies from observation to observation. 
Individual observation values are combined into an average normalisation factor ($\alpha $) quoted in Table \ref{tab:sig}.
Similar excess event counts and significances are observed in both regions, thus attesting to the bipolar morphology of the remnant in the TeV energy range. This is a highly constraining result, because due to the relatively uniform target density around the remnant the H.E.S.S. morphology directly reflects the distribution of high-energy particles responsible for the $\gamma$-ray emission. 

\begin{table}[htbp]
\begin{center}
\begin{tabular}{|c|c|c|c|c|c|}\hline
Region & ON & OFF & $\alpha$ & \# $\gamma$ & Significance \\ \hline
NE, Std Cuts & 4306 & 25421 & 6.67 & 495 & 7.3     \\
NE, Hard Cuts & 619 & 2575 & 6.44 & 219 & 9.3      \\
SW, Std Cuts & 3798 & 26523 & 7.615 & 315 & 4.9     \\
SW, Hard Cuts & 548 & 2591 & 7.25 & 191 & 8.7      \\
\hline
\end{tabular}
\caption{\it H.E.S.S. excess events and significances for the two regions defined from X-ray observations. $\alpha$ is the normalisation factor between OFF and ON exposures.} \label{tab:sig}
\end{center}
\end{table}

\begin{figure}
\begin{center}
\includegraphics[width=0.95\linewidth]{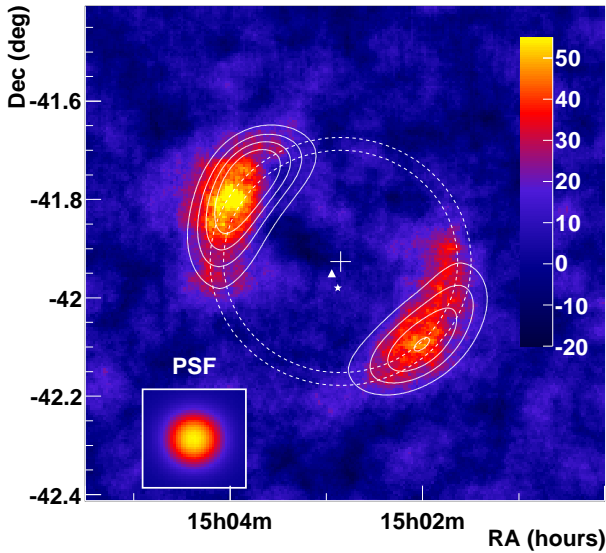}
\caption{\it H.E.S.S. $\gamma$-ray image of SN~1006. The linear colour scale is in units of excess counts per $\pi \times (0.05^\circ)^2$. Points within $(0.05^\circ)^2$ are correlated. The white cross indicates the geometrical centre of the SNR obtained from XMM data as explained in the text and the dashed circles correspond to $R\pm dR$ as derived from the fit. The white star shows the centre of the circle encompassing the whole X-ray emission as derived by Rothenflug et al. (2004) and the white triangle the centre derived by Cassam-Chena\"i et al. (2008) from H$\alpha$ data. The white contours correspond to a constant X-ray intensity as derived from the XMM-Newton flux map and smoothed to the H.E.S.S. point spread function, enclosing respectively  80\% ,  60\% ,  40\%  and  20\% of the X-ray emission. The inset shows the H.E.S.S. PSF using an integration radius of $0.05^\circ$. }
\label{fig:excess200}
\end{center}
\end{figure}

\begin{figure}
\begin{center}
\includegraphics[width=0.95\linewidth]{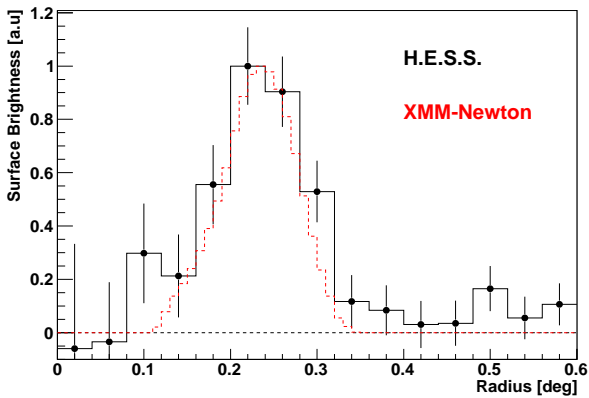}
\caption{\it Radial profile around the centre of the SNR obtained from H.E.S.S. data and XMM-Newton data in the 2 - 4.5 keV energy band smoothed to H.E.S.S. PSF.}
\label{fig:rad}
\end{center}
\end{figure}

\begin{figure}
\begin{center}
\includegraphics[width=0.95\linewidth]{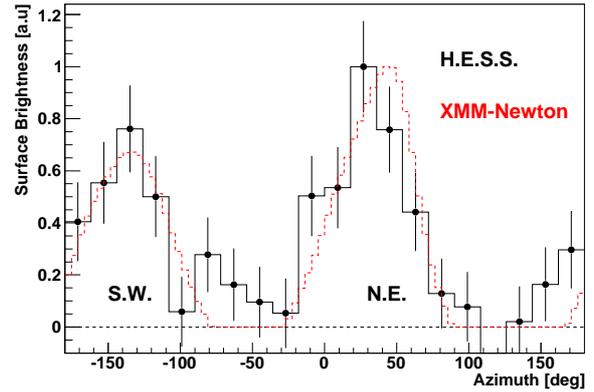}
\caption{\it Azimuthal profile obtained from H.E.S.S. data and XMM-Newton data in the 2 - 4.5 keV energy band and smoothed to H.E.S.S. PSF, restricted to radii $0.12^\circ\leq r\leq 0.36^\circ$ from the centre of the SNR. Azimuth $0^\circ$ corresponds to East, $90^\circ$ corresponds to North, $180^\circ$ to West and $-90^\circ$ to South. }
\label{fig:azi}
\end{center}
\end{figure}

Figure \ref{fig:excess200} shows the $\gamma$-ray H.E.S.S. excess map, produced with hard cuts and the same integration radius of $0.05^\circ$, overlaid with the smoothed XMM-Newton flux contours. A striking similarity between the $\gamma$-ray and X-ray emission regions is found. For a quantitative analysis uncorrelated radial and azimuthal profiles of the H.E.S.S. excess events were derived and compared to the XMM-Newton profiles (Figs. \ref{fig:rad} and \ref{fig:azi}). Again the XMM-Newton data were smoothed to match the H.E.S.S. point spread function, and the relative normalisation was adjusted to the maximum value. Within error bars, the H.E.S.S. and XMM-Newton emission profiles are almost identical, thus possibly indicating a common origin.

The geometrical X-ray centre of the SNR was derived from the unsmoothed XMM-Newton data by fitting 
them with a Gaussian radial profile convolved with an azimuthal profile with two Gaussian components,
yielding \hbox{15h2m51.1s}, \hbox{-41d55'32.2"} as the centre of the SNR with a radius of $R=0.239^{\circ}$ and a thickness of $dR=0.013^{\circ}$. 
Figure \ref{fig:rad} shows the radial profiles of H.E.S.S. and smoothed XMM-Newton excess events from the centre of the SNR. When a Gaussian is fit to the H.E.S.S.~profile (Fig.~\ref{fig:rad}) the shell radius is found to be $0.24^{\circ}\pm 0.01^{\circ}$ and the width of the radial distribution is $0.05^{\circ}\pm 0.01^{\circ}$, which is consistent with the H.E.S.S.~point spread function, thereby showing that the emission region is compatible with a thin rim. 

The azimuthal profile, restricted to radii $0.12^\circ\leq r\leq 0.36^\circ$ from the centre of the SNR, is shown in Fig. \ref{fig:azi} for  H.E.S.S. data and smoothed XMM-Newton data in the 2 - 4.5 keV energy band. The azimuth is defined clockwise
with zero toward the East. The H.E.S.S. profile is compatible with a superposition of two Gaussian emission regions almost at $180^\circ$ from each other,
respectively centred on $-143.6^\circ\pm 6.1^\circ$ (SW region) and $29.3^\circ\pm 4.0^\circ$ (NE region) and with similar widths of $ 33.8^\circ\pm 7.0^\circ$and $27.9^\circ\pm 4.0^\circ$.


\section{Spectral analysis}\label{spectra}

Differential energy spectra of the VHE $\gamma$-ray emission were derived for both regions above the energy threshold of $\sim$  260~GeV. These regions correspond
to 80\% of the X-ray emission (after smearing with the H.E.S.S. PSF)
and therefore slightly underestimate the TeV emission of the full remnant.

\begin{figure}
\begin{center}
\includegraphics[width=0.95\linewidth]{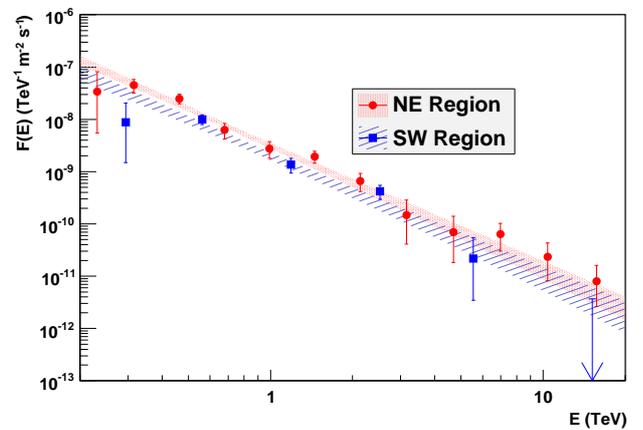}
\caption{\it Differential energy spectra of SN~1006 extracted from the two regions NE and SW as defined in Sect.~\ref{analysis}. The shaded bands correspond to the range of the power-law fit, taking into account statistical errors. }
\label{fig:spectra}
\end{center}
\end{figure}

The spectra for the NE and SW regions are compatible with power law distributions, $F(E) \propto E^{-\Gamma}$, with  comparable photon indices $\Gamma$
and  fluxes. Confidence bands for power-law fits are shown in Fig. \ref{fig:spectra} and
Table \ref{tab:spectra}. The integral fluxes above 1~TeV correspond to less than $1\%$ of the Crab flux, making SN~1006
one of the faintest known VHE sources (Table \ref{tab:spectra}). The derived fluxes are well below the previously published
H.E.S.S. upper limits \cite{HESS_limit}. The observed photon index $\Gamma \approx 2.3$ is somewhat steeper than generally expected from diffusive shock acceleration theory and may indicate that the upper cut-off of the high-energy particle distribution is being observed; however, the uncertainties on the spectrum preclude definitive conclusions on this point. 

\begin{table}[htbp]
\begin{center}
\begin{tabular}{|c|c|c|}\hline
Region & photon index $\Gamma$ & $\Phi(>1 \mathrm{TeV})$ \\ 
& & $(10^{-12} \mathrm{cm}^{-2}\mathrm{s}^{-1})$  \\ \hline 
NE & $2.35 \pm 0.14_{stat} \pm 0.2_{syst}$  & $0.233 \pm 0.043_{stat} \pm 0.047_{syst}$ \\
SW & $2.29 \pm 0.18_{stat}\pm 0.2_{syst}$ & $0.155 \pm 0.037_{stat} \pm 0.031_{syst} $ \\

\hline
\end{tabular}
\caption{\it Fit results for power-law fits to the energy spectra. } \label{tab:spectra}
\end{center}
\end{table}


\section{Discussion}\label{discussion}

The source SN 1006 is an ideal example of a shell-type supernova remnant because it represents a type Ia supernova exploding into an approximately uniform medium and magnetic field, thereby essentially maintaining the spherical geometry of a point explosion. This can be attributed to the fact that SN~1006 is about 500~pc above the Galactic plane in a relatively clean environment, where the external gas density is rather low, $\mathrm{n_H} \approx$ 0.085 cm$^{-3}$ as indicated by Katsuda et al. (2009). Moreover, SN~1006 is one of the best-observed SNRs with a rich data-set of astronomical multi-wavelength information in radio, optical and X-rays, and all the important parameters like the ejected mass, its distance and age are fairly well-known \cite{Cassam}. For this reason, the semi-analytical models of Truelove $\&$ McKee (1999) can be approximately applied and the velocity of the shock calculated. The value of the shock velocity calculated by this means agrees well with the recent measurement in X-rays by Katsuda et al. (2009), yielding $(0.48 \pm 0.04)$~arcsec~yr$^{-1}$ in the synchrotron emitting regions (NE and SW), which corresponds to $5000 \pm 400$~km/s for a distance of 2.2~kpc. This does not contradict the value of $(0.28 \pm 0.008)$~arcsec~yr$^{-1}$ measured by Winkler et~al.~(2003) in the optical filaments, which are situated in the NW region of the remnant. All those calculations neglect the dynamic role of accelerated particles however, which is potentially quite important. 

The basic model of VHE $\gamma$-ray production requires particles accelerated to multi-TeV energies and a target comprising photons and/or matter of sufficient density. The close correlation between X-ray and VHE-emission points toward particle acceleration in the strong shocks revealed by the Chandra observation of the X-ray filaments. Moreover, the bipolar morphology of the VHE-emission in the NE and SW regions of the remnant supports a major result of diffusive shock acceleration theory, according to which efficient injection of suprathermal downstream charged nuclear ions is only possible for sufficiently small angles between the ambient magnetic field and shock normal, and therefore a higher density of accelerated nuclei at the poles is predicted \cite{Ellison1995,Malkov95,Voelk2003}. 

Radio \cite{Reynolds} and  X-ray \cite{Bamba2008} data integrated over the full remnant were combined with VHE $\gamma$-ray measurements to model the spectral energy distribution of the source in a simple one-zone stationary model. 
For the sake of consistency, the VHE $\gamma$-ray energy distribution was determined from the sum of the two previously defined regions. 
In this phenomenological model the current distribution of particles (electrons and/or protons) is prescribed with a given spectral shape corresponding to a power law with an exponential cutoff,
from which emission due to synchrotron radiation, bremsstrahlung and IC scattering on the Cosmic Microwave Background (CMB) photons is computed. The $\pi^0$ production through interactions of protons with the
ambient matter are obtained following Kelner et~al.~(2006). 

 It is clear that this model oversimplifies the acceleration process in an expanding remnant, as discussed by e.g. Drury et~al.~(1989) and Berezhko et~al.~(1996). In addition one must include the uncertainties introduced by the dynamics of the ejecta, the nonuniform structure of the ambient medium and the complexities of the reaction of the accelerated particles on both the magnetic field and the remnant dynamics. This is of importance when comparing the data to the model results below. 

\begin{figure}
\begin{center}
\includegraphics[width=0.95\linewidth]{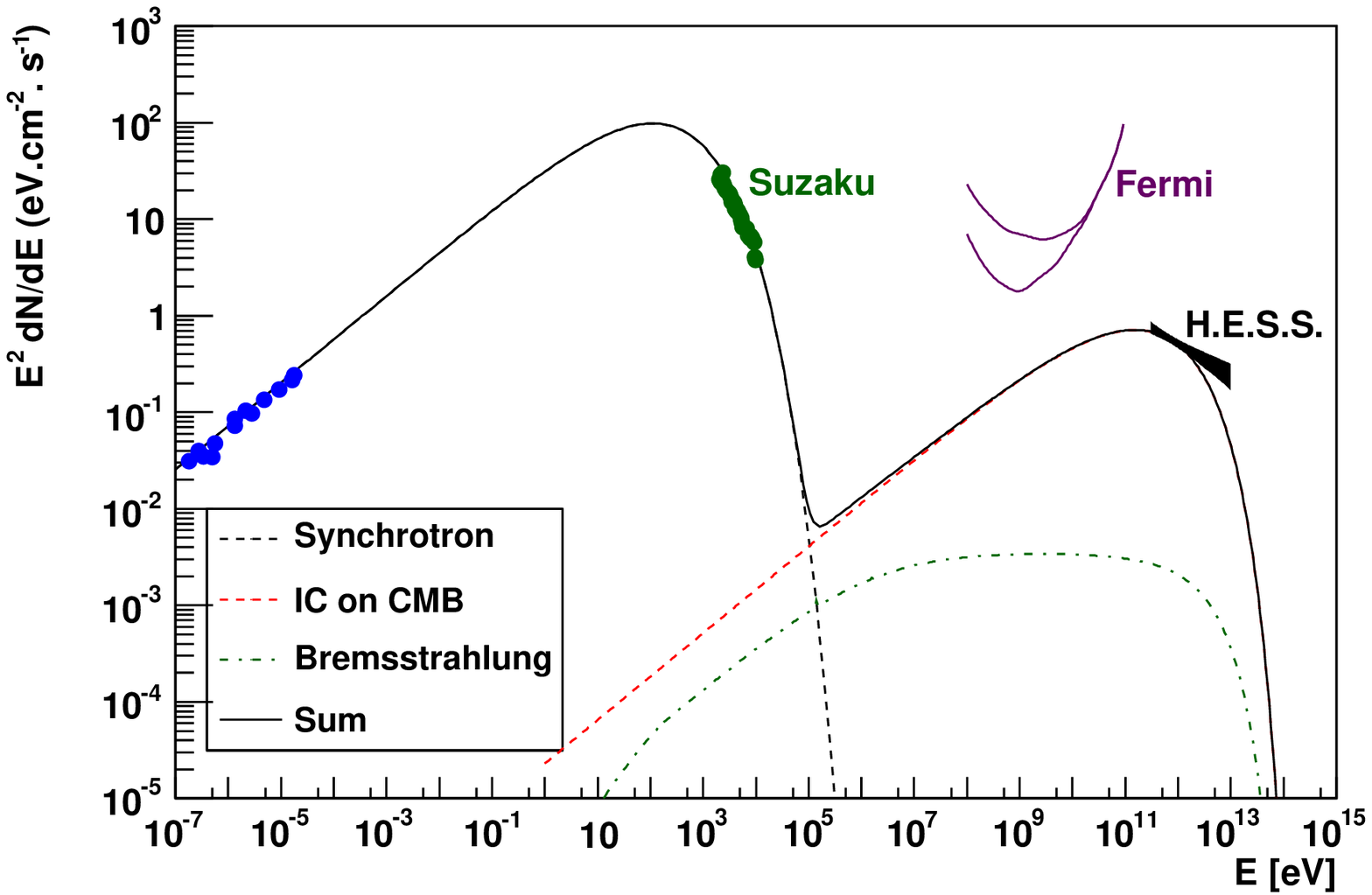}
\includegraphics[width=0.95\linewidth]{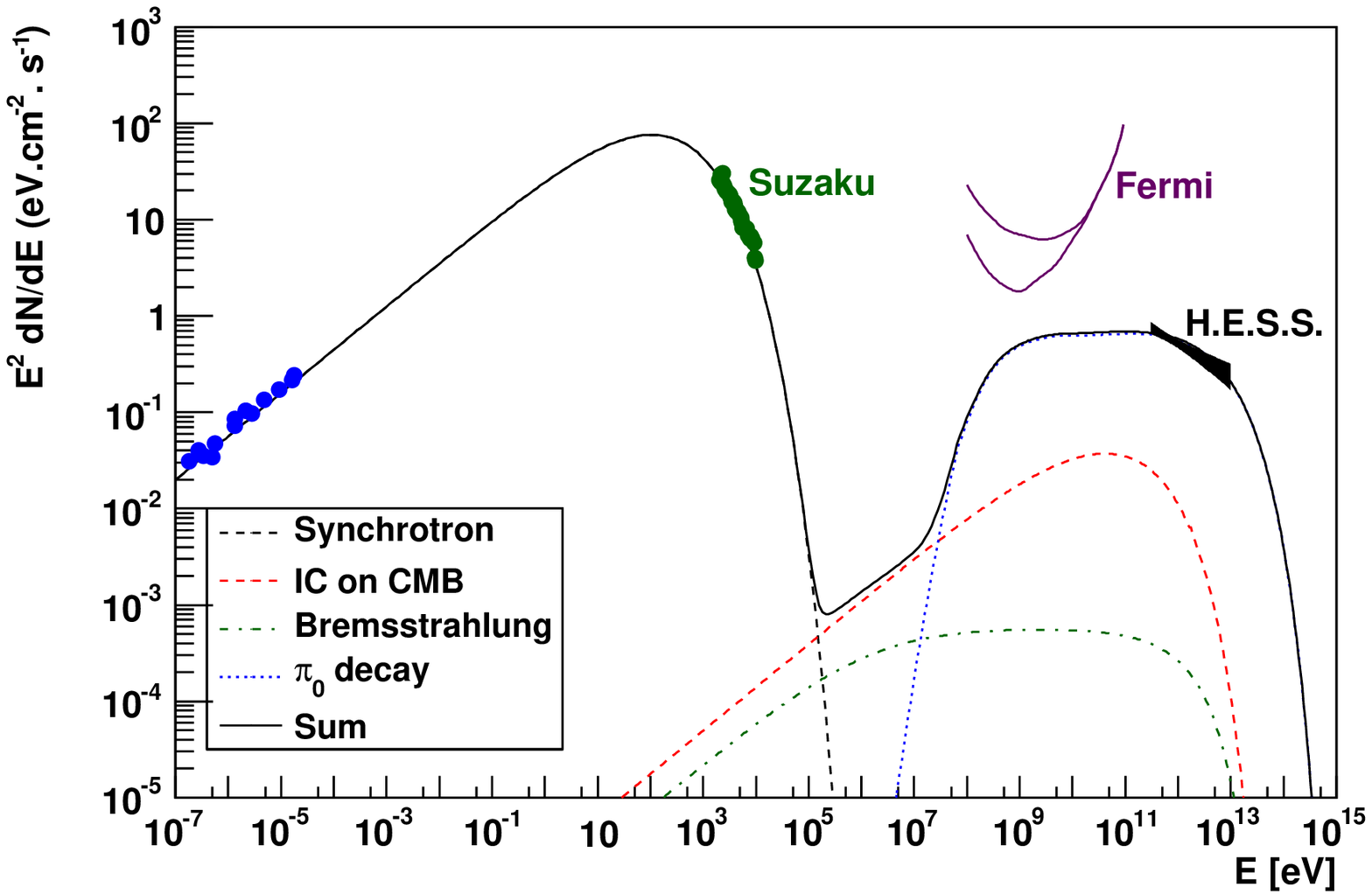}
\includegraphics[width=0.95\linewidth]{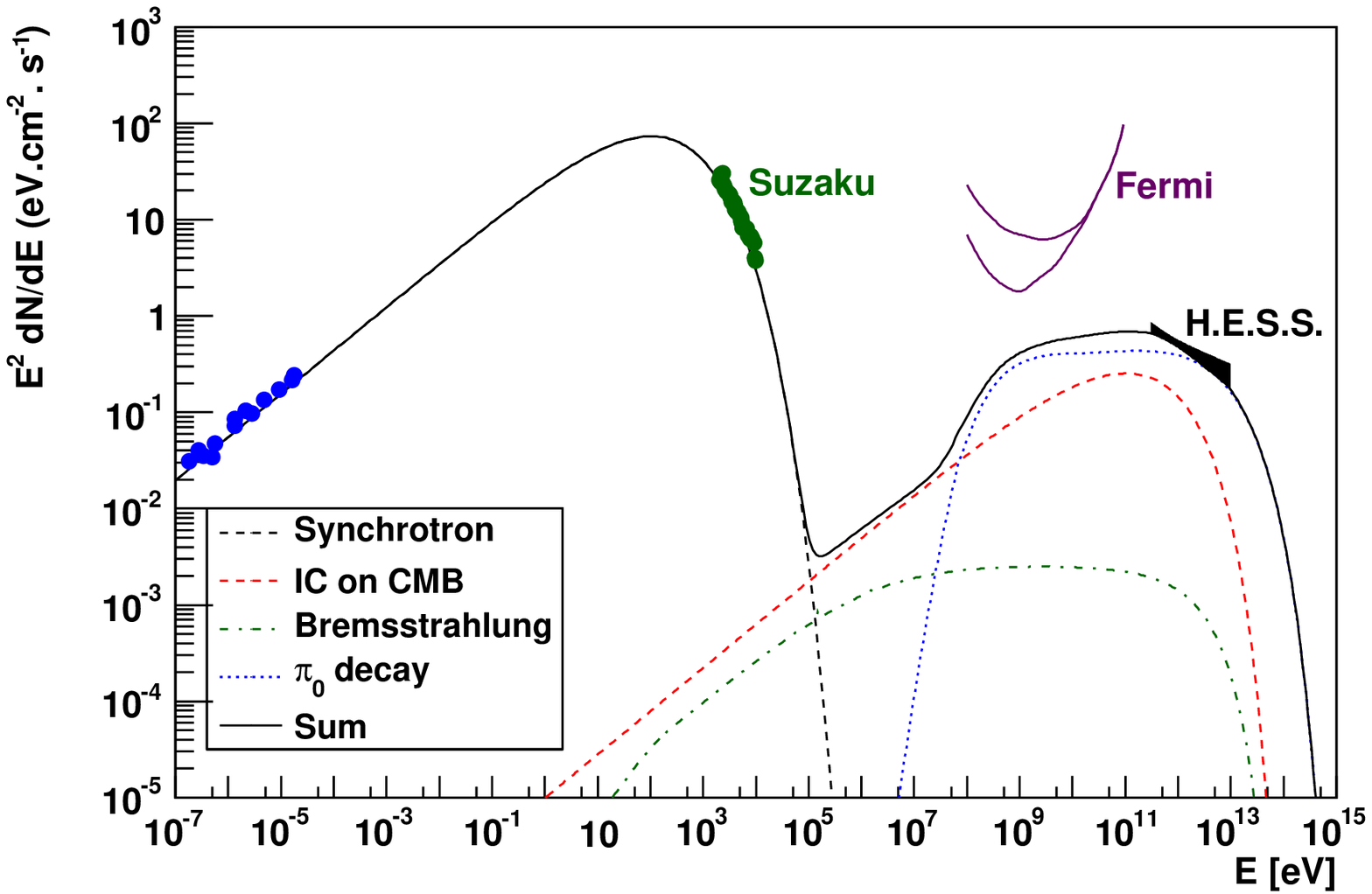}
\caption{\it Broadband SED models of SN~1006 for a leptonic scenario (top), a hadronic one (centre) and a mixed leptonic/hadronic scenario (bottom). 
Top: Modelling was done by using an electron spectrum in the form of a power law with an index of 2.1, an exponential cutoff at 10~TeV and a total energy of $\mathrm{W_e} = 3.3 \times 10^{47}$ erg. The magnetic field amounts to 30~\hbox{\textmu}G. 
Centre: Modelling using a proton spectrum in the form of a power law with an index of 2.0, an exponential cutoff at 80~TeV and a total proton energy of $\mathrm{W_p} = 3.0 \times 10^{50}$~erg (using a lower energy cut off of 1 GeV).
The electron/proton ratio above 1~GeV was $\mathrm{K_{ep}} =  1 \times 10^{-4}$ with an electron spectral index of 2.1 and cutoff energy at 5~TeV. The magnetic field amounts to 120~\hbox{\textmu}G and the average medium density is 0.085~cm$^{-3}$.  
Bottom: Modelling using a mixture of the above two cases.  The total proton energy was $\mathrm{W_p} = 2.0 \times 10^{50}$~erg, with $\mathrm{K_{ep}} =  7 \times 10^{-3}$, with exponential cutoffs at 8~TeV and 100~TeV for electrons and protons respectively. The magnetic field amounts to 45~\hbox{\textmu}G.
The radio data \cite{Reynolds}, X-ray data \cite{Bamba2008} and H.E.S.S. data (sum of the two regions) are indicated. The following processes have been taken into account: synchrotron radiation from primary electrons (dashed black lines), IC scattering (dotted red lines), bremsstrahlung (dot-dashed green lines) and proton-proton interactions (dotted blue lines). The Fermi/LAT sensitivity for one year is shown (pink) for Galactic (upper) and extragalactic (lower) background. The latter is more representative given that SN 1006 is $14^\circ$ north of the Galactic plane. }
\label{fig:sed}
\end{center}
\end{figure}

Assuming first a purely leptonic form
(Fig. \ref{fig:sed}, top), the radio and X-ray data constrain the synchrotron part of the SED in a way that the slope of the electron spectrum, which is particularly sensitive to the slope of the radio data, is bounded between 2.0 and 2.2, while the cutoff energy of electrons is limited to about 10~TeV by the X-ray data assuming a magnetic field of 30~\hbox{\textmu}G. 
With the particle spectrum constrained by radio and X-ray data, the resulting magnetic field needs to be higher than 30~\hbox{\textmu}G so that the IC emission does not exceed the measured VHE-flux. A magnetic field of 30~\hbox{\textmu}G implies that assuming Bohm diffusion, electrons of 1~TeV are confined in a shell of  the width of 10 arcseconds, which is much smaller than the PSF of the H.E.S.S. instrument and is therefore compatible with the radial profile shown in Fig. \ref{fig:rad}. However, while this simple leptonic scenario can account for the measured VHE $\gamma$-ray flux, it fails to reproduce the slope of the VHE spectrum, which is much harder than the expectations from the IC process (see Fig. \ref{fig:sed} top). But it should be noted that non-linear Fermi shock acceleration as reviewed by Malkov  \& Drury~(2001) usually predicts curved cosmic ray spectra with different spectral shapes for protons and electrons. There is a hint of spectral curvature observed in the case of Tycho's and Kepler's supernova remnants in the radio regime \cite{Reynolds&Ellison}. For SN 1006 there is also an indication of the curvature of the electron spectrum in the GeV to TeV energy range \cite{Allen}. These non-linear effects, which also might well introduce a spectral curvature in the VHE regime, are not addressed by this simple model.

In a second dominantly hadronic model (Fig. \ref{fig:sed}, middle) TeV emission results from proton-proton interactions 
with $\pi^0$-production and subsequent decay, 
whereas the X-ray emission is still produced by leptonic interactions. A rough representation of the effect of spectral curvature is included by allowing for a slightly harder spectral index for protons than for radio-emitting electrons.
A lower electron fraction allows us to account for the X-ray and radio emission with a higher field value of 120~\hbox{\textmu}G, which is consistent with magnetic field amplification at the shock, as indicated by the above-mentioned measurements of thin X-ray filaments.
Assuming an average medium density of 0.085~cm$^{-3}$ and a proton spectral index of 2.0 with a cutoff energy of 80~TeV (inferred from the maximum energy of TeV photons), this model requires a high overall fraction of about 20\%, of the supernova energy to be converted into high-energy protons. 
Here $E_{SN} = 1.4 \times 10^{51}$~erg was assumed, near the upper end of the typical range of type Ia SN explosion energies (e.g.\ Woosley et al.\ 2007), as the assumed density, observed radius and known age of SN 1006 appear to require a higher than average explosion energy. Given that the VHE emission is concentrated in polar regions of the shell, the local shock acceleration efficiency would then be several times higher than this fraction.

In a third example (mixed model), hadronic and leptonic processes contribute almost equally to the very high-energy emission.
The electron spectrum is similar to the aforementioned leptonic case and the total proton energy is set to 14\% of the mechanical supernova energy with the electron/proton ratio $K_\mathrm{ep} = 3.9 \times 10^{-3}$, thus leaving the magnetic field and the cutoff energy of protons the only free parameters.
In the example shown in Fig. \ref{fig:sed} (bottom panel) the magnetic field amounts to 45~\hbox{\textmu}G and the cutoff energy of protons is 100~TeV. This example illustrates that in this simple one-zone case it is possible to reproduce all the multi-wavelength data on SN~1006 to a reasonable degree of accuracy including the slope of the VHE-data. While these considerations cannot exclude any of the astrophysical scenarios, they serve as a 
quantitative illustration of the various alternatives. 

Values of total electron and proton energy, cutoff energy and magnetic field obtained in the three aforementioned cases are summarised in Table \ref{tab:model}. These parameters yield very similar values when the NE and SW regions are adjusted independently.

\begin{table}[htbp]
\begin{center}
\begin{tabular}{|c|c|c|c|c|c|}\hline
Model & $E_{cut,e}$ & $E_{cut,p}$ & $W_e$ & $W_p$ & $B$\\ 
& $[\mathrm{TeV}]$ & $[\mathrm{TeV}]$ & $[10^{47} \mathrm{erg}]$ & $[10^{50} \mathrm{erg}]$ & $[\mathrm{\mu G}]$ \\\hline
Leptonic & $10$ & -  & $3.3$ & - & $30$ \\
Hadronic & $5$ & $80$ & $0.3$ & $3.0$ & $120$ \\
Mixed & $8$ & $100$ & $1.4$ & $2.0$  & $45$ \\ 
\hline
\end{tabular}
\caption{\it Parameters used in the spectral energy modelling shown in Fig.~\ref{fig:sed}. Spectral indices have been fixed  to
$2.1$ and $2.0$ respectively for electrons and protons.} \label{tab:model}
\end{center}
\end{table}

More elaborate models using e.g. a nonlinear kinetic acceleration theory \cite{Heinz} go beyond the simple approach developed here and lead to precise predictions that could
be quantitatively tested against the data. Several effects which were not included in the simple model above would alter the total energy in accelerated particles required for the hadronic component.  Beyond the spectral curvature mentioned previously, these include the higher compression of the target matter induced by the dynamical reaction of the accelerated particles, and consideration of the heavier nuclei composition of the accelerated hadrons instead of the pure protons assumed here.
Measurements in the GeV-energy range would be pivotal to distinguish between the different scenarios.
Unfortunately, the sensitivity of the Fermi Large Area Telescope for one year as given in Atwood et~al.~(2009) is of a factor of the order of 10 too low (depending on the model and the exact diffuse background flux) to measure the predicted flux at 1~GeV as shown in Fig.\ref{fig:sed}, which makes the detection of SN 1006 by Fermi LAT rather unlikely.


\section{Conclusions}\label{conclusion}

Very high energy $\gamma$-rays from SN 1006 have been detected by H.E.S.S. The measured flux above 1~TeV is of the order of 1$\%$ of that detected from the Crab Nebula and therefore compatible with the previously published upper limit \cite{HESS_limit}. The bipolar morphology apparent in $\gamma$-rays is consistent with the non-thermal emission regions also visible in X-rays. As the VHE-shell is compatible with a scenario of thin rim emission, particle acceleration in the very narrow X-ray filaments, which are signatures of shocks, is also likely to be at the origin of the $\gamma$-ray signal.  The measured flux level can be accounted for by inverse Compton emission assuming a magnetic field of about 30~$\mu$G.  A mixed scenario including leptonic and hadronic processes and taking into account the ambient matter density estimated from observation also leads to a satisfactory description of the multi-wavelength spectrum, assuming a high proton-acceleration efficiency.  None of the models can be excluded at the level of modelling presented here.


\begin{acknowledgements}
The support of the Namibian authorities and of the University of Namibia in facilitating the construction and operation of H.E.S.S. is gratefully acknowledged, as is the support by the German Ministry for Education and Research (BMBF), the Max Planck Society, the French Ministry for Research, the CNRS-IN2P3 and the Astroparticle Interdisciplinary Programme of the CNRS, the U.K. Science and Technology Facilities Council (STFC), the IPNP of the Charles University, the Polish Ministry of Science and Higher Education, the South African Department of Science and Technology and National Research Foundation, and by the University of Namibia. We appreciate the excellent work of the technical support staff in Berlin, Durham, Hamburg, Heidelberg, Palaiseau, Paris, Saclay, and in Namibia in the construction and operation of the equipment.
  
\end{acknowledgements}

\newcommand{\thc}{(\hess\ Collaboration) }

\end{document}